\documentclass[twocolumn,fleqn,natbib]{svjour2}
\bibpunct{[}{]}{;}{n}{}{,} 
\smartqed  
\usepackage{graphicx}
\usepackage{color}

\journalname{Granular Matter}
\begin{document}

\title{Velocity profiles in forced silo discharges}

\titlerunning{Forced silo discharges}        

\author{Marcos A. Madrid \and Luis A. Pugnaloni
} \institute{Marcos A. Madrid: Dpto. Ingenier\'ia Mec\'anica, Facultad Regional La Plata, Universidad Tecnol\'ogica Nacional, CONICET, Av. 60 Esq. 124, 1900 La Plata, Argentina. E-mail: marcosamadrid@gmail.com
               \\
             L. A. Pugnaloni: Dpto. Ingenier\'ia Mec\'anica, Facultad Regional La Plata, Universidad Tecnol\'ogica Nacional, CONICET, Argentina. {\it Current address:} Dpto. F\'isica, FCEyN, Universidad Nacional de La Pampa, Uruguay 151, 6300 Santa Rosa, La Pampa, Argentina. E-mail: luis.pugnaloni@exactas.unlpam.edu.ar
}

\authorrunning{M. A. Madrid et al.} 

\date{Received: date }

\maketitle

\begin{abstract}
 When a granular material is freely discharged from a silo through an orifice at its base, the flow rate remains constant throughout the discharge. However, it has been recently shown that, if the discharge is forced by an overweight, the flow rate increases at the final stages of the discharge, in striking contrast to viscous fluids [Madrid et al. Europhys. Lett. (2018)]. {Although the general mechanism that drives this increase in the flow rate has been discussed, there exist yet a number of open questions regarding this phenomenon. One such questions is to what extent is the internal velocity profile affected,  beyond the trivial overall increase consistent with the increasing flow rate.} We study via Discrete Element Method simulations the internal velocity profiles during forced silo discharges and compare them with those of free discharges. The changes in velocity profiles are somewhat subtle. Interestingly, during free discharges, while the velocity profiles are steady at the silo base and above a height equivalent to one silo diameter, there exists a transition region where the profile evolves in time, despite the constant flow rate. In contrast, forced discharges present steady profiles at all heights of the granular column during the initial constant flow phase, followed by an overall increase of the velocities when the acceleration phase develops.
\end{abstract}

\section{Introduction}\label{sec:into} 

The discharge of granular materials through an orifice at the bottom of a silo is a phenomenon profusely studied (see for example \cite{kadanoff1999,degennes1999,Jaeger1996,duran2000,tighe2007,beverloo,NeddermanBook}). This is in part due to the technological interest of the problem, but also because the significant challenges the scientific community has encountered in predicting basic quantities such as flow rate and wall pressure during discharge. The flow rate can be calculated by using the empirical Beverloo equation \cite{beverloo}. However, there is not a full theoretical derivation of this equation based on first principles. The main features of the Beverloo equation are explained by heuristic postulates such as the \emph{free fall arch} and the \emph{empty annulus} \cite{tighe2007,NeddermanBook}. Bottom pressure and wall pressure during discharge is also poorly understood. It is often assumed that the Janssen equation \cite{janssen,sperl2006} can predict the internal pressure under flowing conditions, despite a number of experiments demonstrating that this is not the case \cite{aguirre2010,wang2015,perge2012} and a few attempts to expand the Janssen approach to flowing silos \cite{walters1973,walters1973b,walker1966,walker1967}. 

A remarkable feature of free granular discharges is the constant flow rate observed despite the drop in the height of the granular column. This is in contrast with viscous fluids, which show a decreasing flow rate as the level of the container falls. It is also surprising that the flow rate seems to be rather insensitive to the details of the grains properties (friction, elastic modulus, poison ration, etc.). However, this is somehow consistent with viscous fluids where the flow rate through an orifice on a plate does not depend on its viscosity \cite{Linford1961}.  

\begin{figure}[t]
	\centering
	\includegraphics[width=0.9\linewidth,trim= 0cm 1.2cm 0 1cm, clip]{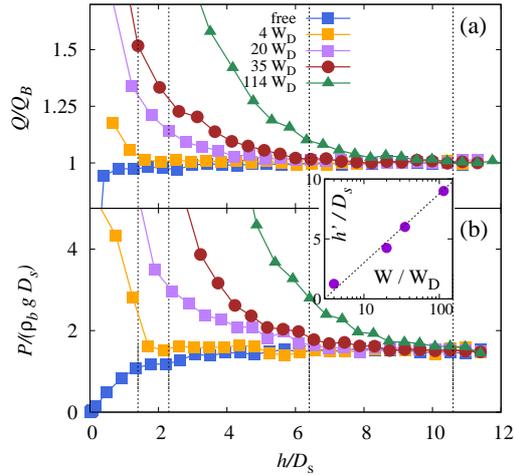}
	\caption{(a) Particle flow rate $Q$ as a function of the column height $h$ for various overweights $W$ (see legend). $Q$ is scaled by the Beverloo flow rate ($Q_{\rm B}$).  (b) Bottom pressure during the discharge. The vertical lines correspond to the different column heights $h_{\rm fill}$ along the discharge for which we have measured the various velocity profiles. Inset: Column height $h'$ at which the acceleration phase starts as a function of the overweight $W$ (note the log-lin scale).}
	\label{fig:q-p-vs-t}
\end{figure}

\begin{figure}[t]
	\centering
	\includegraphics[width=0.99\linewidth,trim=0 2cm 0 0,clip]{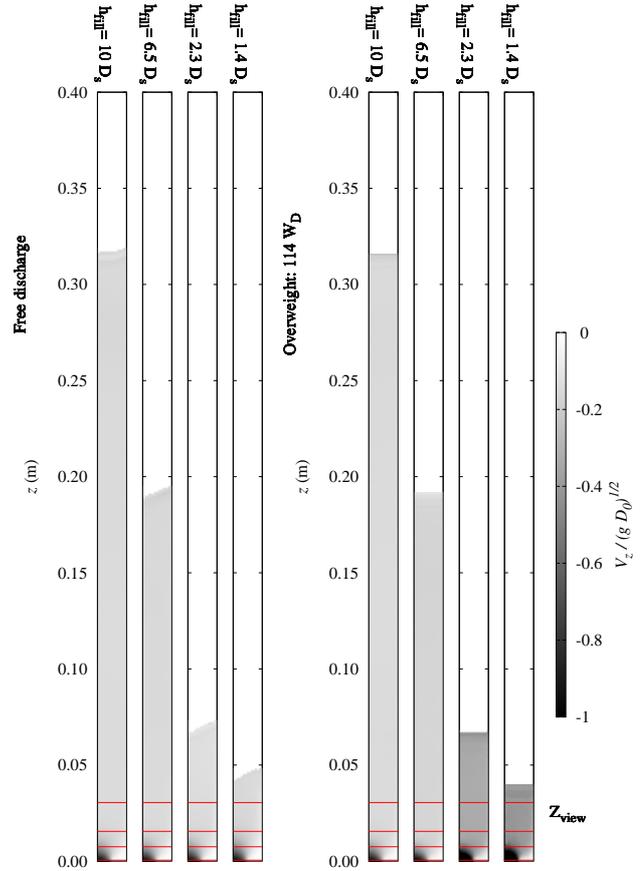}
	\caption{Vertical velocity heat map, $v_z(r,z)$, of the silo at different stages of the discharge (different $h_{\rm fill}$) for a free discharge (left) and for a forced discharge with an overweight of $W =114 W_{\rm D}$ (right). Note that only the right half of the silo is shown. The labels $h_{\rm fill}$ on each figure indicate the point at which the heat map has been extracted during the discharge according to the vertical lines indicated in Fig. \ref{fig:q-p-vs-t}(a). The horizontal red lines indicate the position $z=z_{\rm view}$ at which profiles are measured in the rest of the paper.}
	\label{fig:vy-heatmap}
\end{figure}

Recently, it has been shown that the use of an overweight on top of the granular column of a discharging silo induces an increase of the flow at the final stages of the discharge \cite{madrid2018}. This is once again in contrast to viscous fluids since these decrease their flow rate during discharge from a tank even if a constant overpressure is applied to the free surface. More striking, however, is the fact that the increase in granular flow rate does depend on the grains material properties, which has no parallel effect in the case of viscous fluids. The increase in flow rate using an overweight is indeed an unexpected feature. Common wisdom among specialized scholars indicates that ``... it is pointless to try to extrude a granular material through a converging passage by the application of a stress.'' as stated in page 304 of Ref. \cite{NeddermanBook}. However, experimental observations are against this prediction \cite{madrid2018}.   

{One may expect that the use of an overweight is equivalent to the addition of extra granular material on top of the original granular column. However, any additional granular material will contribute to the dissipation of energy during the discharge due to the nonconservative particle--particle and particle--wall interactions. A solid overweight operates differently since it lays a constant weight on the free surface of the granular column and does not dissipate energy except for the particle--overweight nonconservative contacts.}

{The basic mechanism to explain the increase in flow rate due to an overweight is based on energy balance \cite{madrid2018}. In a dense granular system, the collision rate is very high, which leads to \emph{inelastic collapse}. Therefore, all the power injected by gravity during the discharge of a silo is continuously dissipated. As a consequence, the discharge proceeds at constant velocity since the energy input cannot be converted into kinetic energy. Since the number of collisions per unit time is so large, all the energy is dissipated even if the material properties are varied. This renders the flow rate material independent. When a piston does work on the system at a high rate, the collisions are unable to fully dissipate the energy input, particularly when the number of particles in the column decreases. As a result, the flow rate increases due to the non-dissipated injected power that is converted into kinetic energy.}

{Due to the increase in flow rate, the mean velocity inside the silo increases in the final phase of the discharge of a forced silo. However, one may expect that the internal} {velocity profile is not simply upscaled but changed in a more complex way. This becomes an important piece of information when it comes to applications. Mass flow and funnel flow patterns are desirable/undesirable depending on the applications.} In this work, we focus on the velocity profiles inside a discharging silo and compare free discharges with forced discharges. We find a previously unnoticed behavior where free discharges present unsteady velocity profiles in some regions of the silo, while forced discharges present a more steady profile throughout the silo during the initial stages before the flow rate increases. During the phase of flow rate increase, the forced flows tend to flatter profiles in comparison with free discharges.

\begin{figure*}[t]
	\centering
	\includegraphics[width=0.8\linewidth,trim= 2cm 1.cm 2cm 1cm, clip]{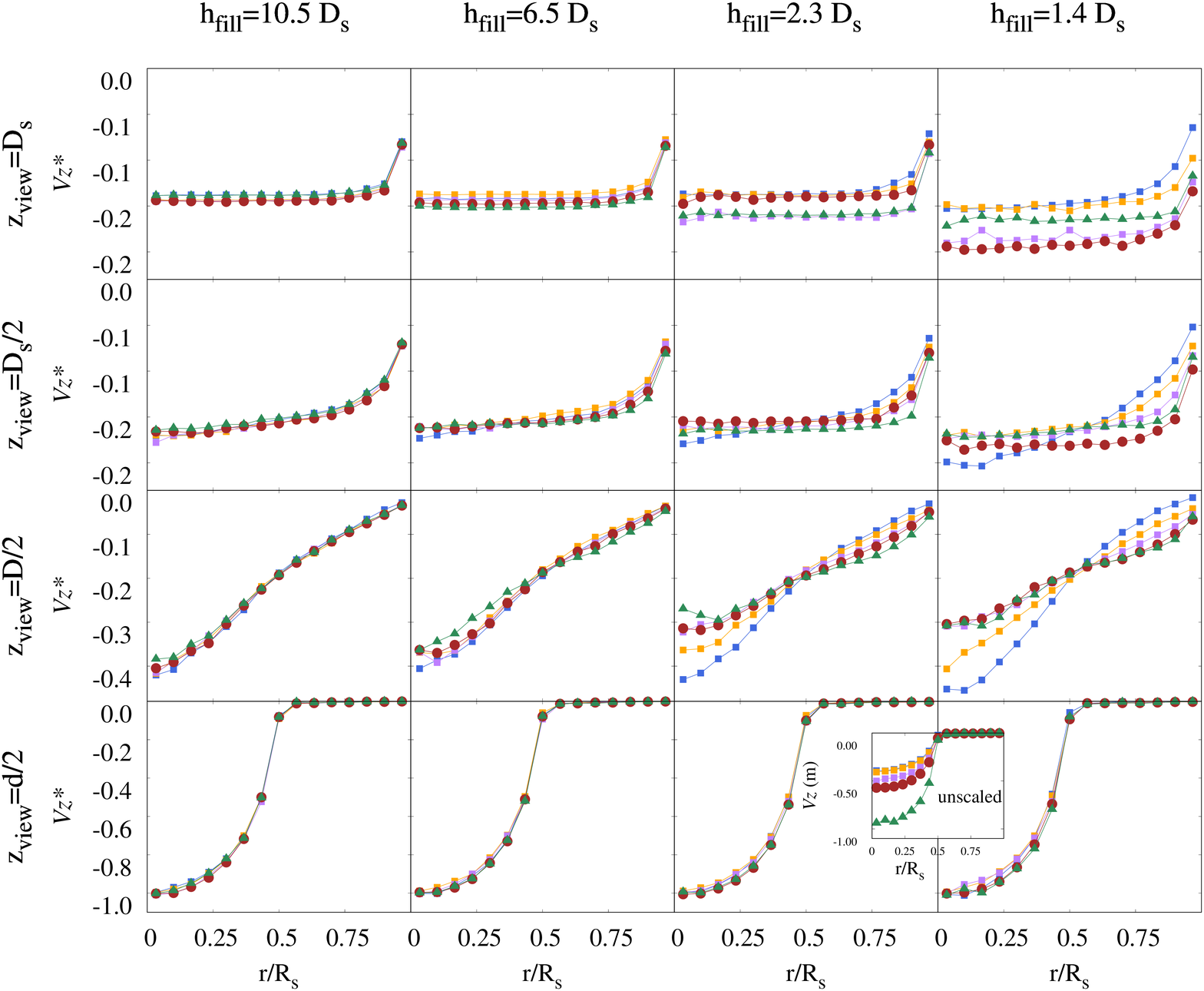}
	\caption{Scaled vertical velocity $v^*_z(r)=v_z(r)/v_z(z=0,r=0)$ at various heights $z_{\rm view}$ (as indicated in Fig. \ref{fig:vy-heatmap}) and for the different column heights $h_{\rm fill}$ during the discharge. Results for different overweights are compared in each plot: free discharge (blue), $W=4W_{\rm D}$ g (orange), $W=20W_{\rm D}$ (purple), $W=35W_{\rm D}$ (red) and $W=114W_{\rm D}$ (green). Each row corresponds to a different observation plane $z_{\rm view}$ in the silo, while each column corresponds to a different column height $h_{\rm fill}$ during the discharge. Inset: Sample of the velocity profile before scaling.}
	\label{fig:vy-vs-r}
\end{figure*}

\section{Simulations}\label{sec:NumSetup}

We use Discrete Element Method (DEM) simulations via the LIGGGHTS \cite{liggghts} implementation with a particle--particle Hertz interaction and Coulomb criterion using a Young modulus $Y=70$ MPa, Poison ratio $\nu=0.25$, restitution coefficient $e=0.95$ and friction coefficient $\mu = 0.5$ \cite{liggghts}. The same interaction applies for the particle--wall contacts.  Particles are spherical with diameter $d=1$ mm and density $\rho=2500$ kg/m$^3$. {The silo is formed by a cylinder of diameter $D_{\rm s}=2R_{\rm s}=30$ mm. The base of the silo is flat and has a circular orifice of diameter $D=15$ mm in its center.}  For the forced flow simulations, we introduce as an overweight a cylindrical piston with the desired weight $W$. Particles are poured to fill a height in the silo above $12 D_{\rm s}$ (which implies up to $3 \times 10^5$ grains). {The bulk volume fraction obtained is $\rho_{\rm b}=1500$ kg/m$^3$}. The orifice is initially blocked by a plug. After the grains come to rest in the silo (we wait until the kinetic energy per particle falls below $5 \times 10^{-12}$ J), we remove the plug and allow its discharge. The acceleration of gravity is $g=9.81$ m/s$^{2}$ and the time step $5\times10^{-6}$ s. {We measure the overweight in terms of the typical weight $W_{\rm D}$ of a column of grains of height $D_{\rm s}$, i.e., $W_{\rm D}=2 \pi R_{\rm s}^3 \rho_{\rm b}$.}

To assess the velocity and pressure profiles at different stages of the discharge we save snapshots with the particle positions, velocities and contact forces every $5\times10^{3}$ time steps (i.e., $40$ fps). We average properties over snapshots grouped by the number of grains that remain in the silo (in blocks of $1\times10^{4}$ particles). This allows us to compare discharges having different flow rates (free and forced discharges) by putting side by side data that correspond to the same average column height. A square grid of side $d$ over the $z-r$ plane is used to create profiles, which are averaged over the azimuth $\theta$ taking advantage of the cylindrical symmetry. The averaged velocity and pressure for each bin in the grid is calculated over the particles in the bin for each snapshot in the desired stage of the discharge. To focus on a particular position in the silo we take slices (usually horizontal slices) of the grid.

\section{Results}\label{sec:Results}

\subsection{Flow rate and bottom pressure}\label{sec:Results-0}

In Fig. \ref{fig:q-p-vs-t}(a), we show the particle flow rate as a function of column height for a free discharging silo along with forced flows using different overweights. As we can see, in the initial stages of a forced discharge, the flow rate is the same as the free flow rate. However, after this initial steady state, an accelerated flow is observed for all forced discharges. The heavier the overweight, the sooner the acceleration phase starts and the faster is the growth of the flow rate. This is consistent with the experimental results shown in Ref. \cite{madrid2018}. {The inset to Fig. \ref{fig:q-p-vs-t} shows that the column height $h'$, at which the accelerated flow starts, grows logarithmically with the overweight $W$. Currently, we can not provide a plausible explanation for this.} As we will see in the next section, despite observing the same flow rate during the initial phase of the discharge, the forced flows display some interesting subtle differences when compared against unforced discharges.  

The pressure at the bottom of the silo during free and forced discharges is shown in Fig. \ref{fig:q-p-vs-t}(b). While the pressure drops quickly at the end of the discharge in an unforced flow, it rises for the forced flows. There seems to be an apparent direct relation between bottom pressure and the flow rate. However, this is not the case. The same particle flow rate can be observed for very different bottom pressures by simply using denser materials and/or wider silos \cite{madrid2018,staron2012}. Also, the bottom pressure increases linearly with gravity, yet the flow rate increases with $g^{0.5}$ \cite{dorbolo2013,arevalo2014}. A discussion on the relation between pressure, flow rate and dissipated power can be found in Ref. \cite{madrid2018}. In this paper we focus only on velocity profiles and leave the discussion on pressure profiles for a future work.

\subsection{Velocity profiles}\label{sec:Results-I}

\begin{figure}[t]
	\centering
	\includegraphics[width=0.9\linewidth]{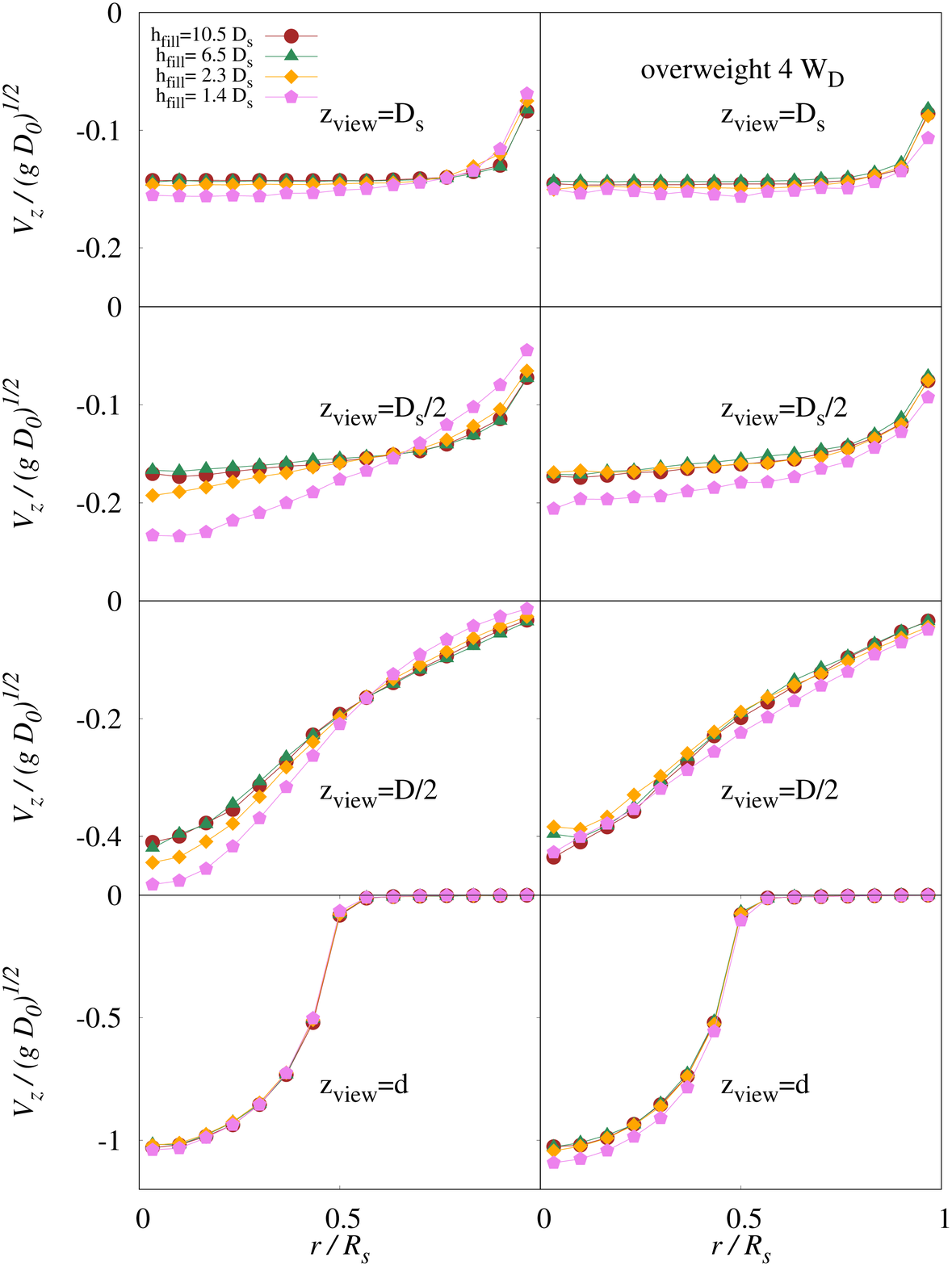}
	\caption{Vertical velocity $v_z(r)$ at various heights $z_{\rm view}$ (as indicated in Fig. \ref{fig:vy-heatmap}) for the free discharge (left column) and for the forced discharge with an overweight of $W=4 W_{\rm D}$ (right column). The different curves represent different column heights $h_{\rm fill}$ during the discharge (see legend).}
	\label{fig:vy-vt-t}
\end{figure}

\begin{figure}[t]
	\centering
	\includegraphics[width=0.9\linewidth]{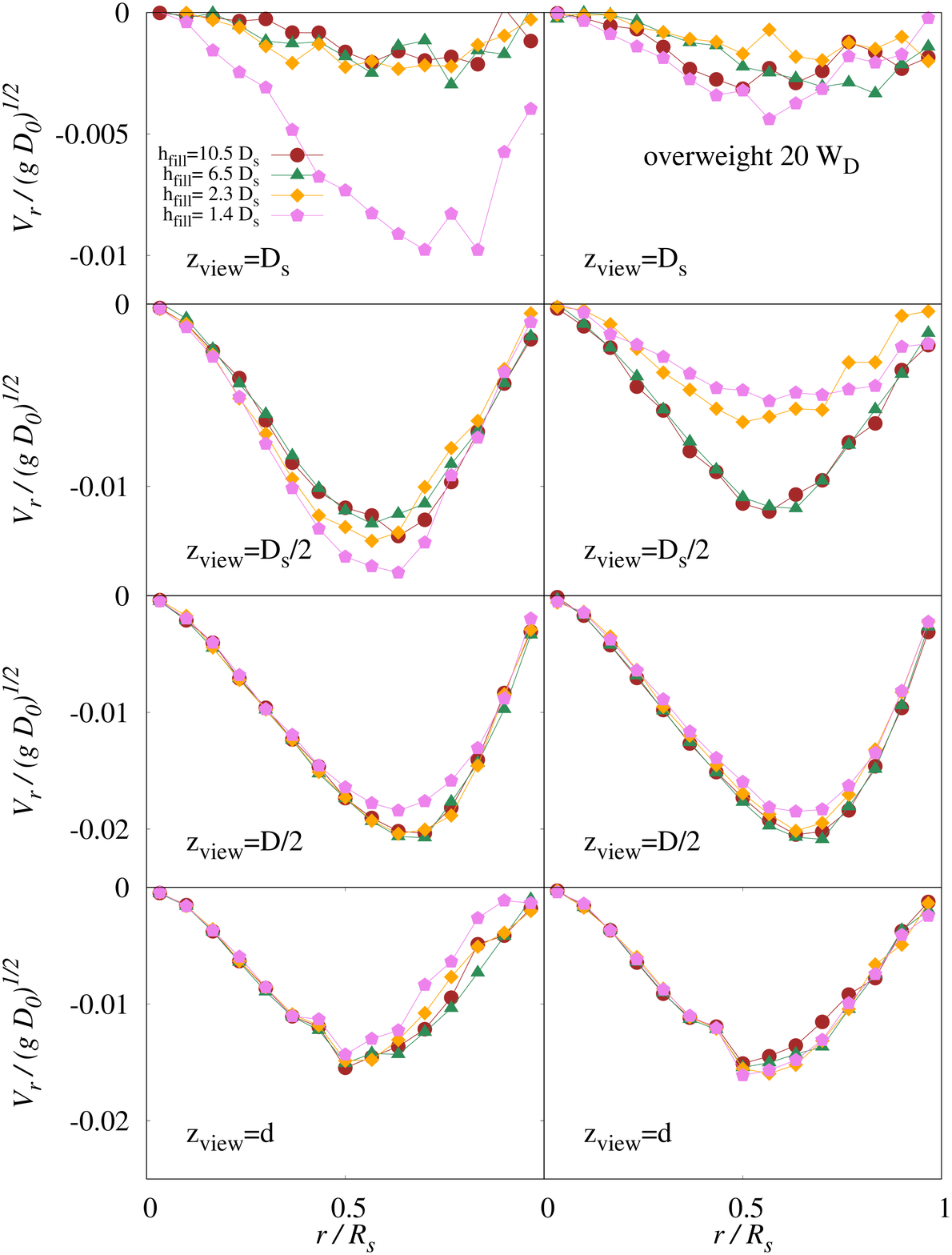}
	\caption{Radial velocity profiles $v_r(r)$  at various observatin planes $z_{\rm view}$ (as indicated in Fig. \ref{fig:vy-heatmap}) for the free discharge (left column) and for the forced discharge with an overweight of $W=20W_{\rm D}$ (right column). The different curves represent different column height $h_{\rm fill}$  during the discharge (see legend).}
	\label{fig:vr-vs-r}
\end{figure}

In Fig. \ref{fig:q-p-vs-t}, we have indicated four different column heights $h_{\rm fill}$ during the discharge (see vertical lines) where we have focused our attention. We compare velocity profiles for different discharges at these different points by ensuring that the granular column has a similar number of grains inside the silo. The corresponding vertical velocity heat maps, $v_z(r,z)$, are shown in Fig. \ref{fig:vy-heatmap} for a free discharge and for a forced discharge with the heaviest overweight studied. 
For the free discharge, we observe the usual overall features described in the literature: (a) there exists a stagnant zone at the base surrounding the orifice ($v_z(r>D/2,z<\alpha r) \approx 0$, with $\alpha$ a constant), (b) there is a focused high velocity region at and above the orifice ($r<D/2, \, z<L\approx D/2$), (c) the free surface of the granular column develops a depression at the center, and (d) the velocity is higher at the central axis of the silo than at the walls ($v_z(r=0)>v_z(r=D_{\rm s}/2)$). Some of these features are hard to see in the scale used in Fig. \ref{fig:vy-heatmap}. The main contrasts shown by the forced discharge are: (a) the stagnant zone is shallower ($\alpha_{\rm forced} < \alpha_{\rm free}$), (b) the high velocity region above the orifice is lower in height ($L_{\rm forced} < L_{\rm free}$), and (c) the overweight forces a flat top surface of the granular column. Of course, we also see the overall increase of the velocities at the final stages where the flow increases as shown in Fig. \ref{fig:q-p-vs-t}(a).

To enhance the contrast between the forced and unforced velocity profiles, we have plotted in Fig. \ref{fig:vy-vs-r} horizontal sections of the profiles (i.e., $v_z(r)$) at different heights $z_{\rm view}=d/2$, $D/2$, $D_{\rm s}/2$ and $D_{\rm s}$ in the silo and for the four column heights $h_{\rm fill}$ considered during the discharge. We have indicated these sections as horizontal lines in Fig. \ref{fig:vy-heatmap}. Moreover, the velocities have been scaled at each point during the discharge (each column) by the velocity at the center of the orifice (notice that for forced flows this velocity increases during discharge).

If we focus on the initial stage of the discharge (left column in Fig. \ref{fig:vy-vs-r}), the scaled velocity profiles at all heights $z_{\rm view}$ and for all the overweights coincide. Actually, in this initial stage of the discharge even the unscaled profiles match. At $z_{\rm view}=D_{\rm s}$ (top row), $v_z(r)$ is rather flat, with a slightly lower velocity next to the walls. However, further down into the granular column the profile becomes steeper. Finally, next to the plane of the base ($z_{\rm view}=d/2$) $v_z$ drops to zero everywhere except at the orifice where the velocity displays a circular profile consistent with Refs. \cite{janda2012,rubio2015}.    

Figure \ref{fig:vy-vs-r} shows that, when the flow rate increases for the forced discharges (two rightmost columns), the velocity profile displays some differences with the free discharge. The most evident feature is an increase in the velocity at the orifice (this can be appreciated in the unsacaled profiles of the inset, bottom--right). However, the shape of the profile when scaled is conserved at the aperture. The flat profile at $z_{\rm view}=D_{\rm s}$ also remains flat upon acceleration. Nevertheless, the region $D/2 < z < D_{\rm s}/2$ presents nonlinear changes in the profile. In general, the scaled profile becomes less steep as the overweight is heavier, reducing $v_z$ at $r=0$ and increasing its value at $r=D_{\rm s}/2$.

%

In Fig. \ref{fig:vy-vt-t}, we show some profiles from Fig. \ref{fig:vy-vs-r} but comparing in each plot different points during the discharge for two forcing condition ($W=0$ and $W=4 W_{\rm D}$). A peculiar feature emerges in these plots. Whereas the velocity profile for the unforced discharge remains unaltered during discharge at the orifice and at a high plane ($z_{\rm view}=D_{\rm s}$), it does evolve in the ``transition region'' $D/2 < z < D_{\rm s}/2$. In this region, $v_z(r=0)$ increases (and $v_z(r=D_{\rm s}/2)$ decreases) as the silo discharges. This is somewhat unexpected since the flow rate is steady and one may assume that the internal velocity profiles are steady too. It is worth mentioning that previous experimental studies have pointed out this velocity drift of the ``plug zone'' \cite{sielamowicz2005,waters2000}. Interestingly, for forced flows, in the transition region $D/2 < z < D_{\rm s}/2$, the profile $v_z(r)$ does remain unaltered during discharge until the start of the accelerated phase of the flow. This is a remarkable feature and implies that one could homogenize the velocity profile in a silo by using an overweight. Of course, in the final stages of the discharge this would cause an increase of the flow rate.    

Some profiles for the radial velocity in the silo are shown in Fig. \ref{fig:vr-vs-r}. The values of $v_r$ are at least one order of magnitude lower than those of $v_z$. In all cases, the radial velocity is zero both at the walls ($v_r(r=D_{\rm s}/2)=0$) and at the axis of the silo ($v_r(r=0)=0$). A maximum in $v_r$ is always found at an intermediate value of $r$. In particular, for $z_{\rm view}=d$ the maximum occurs at $r=D/2$ as expected. Once again, a forced flow yields profiles at all heights that are steady and do not change much with time until the accelerated phase develops. However, the unforced discharge displays increasing radial velocities at $z_{\rm view}=D_{\rm s}/2$, while radial velocities drop during discharge at $z_{\rm view}=D/2$ and $z_{\rm view}=d$. One interesting feature is that for the forced flow, at $z_{\rm view}=D_{\rm s}/2$, the radial velocity decreases significantly during he accelerated phase. Beyond this, overall, the radial velocity is not affected by the fact that the forcing induces an acceleration in the flow rate.

\section{Conclusions}\label{Sec:Concl}

We have shown that the velocity profiles during the discharge of a silo can be affected {in a non trivial way} by the forcing of the flow via an overweight. Since the flow rate increases during the discharge due to forcing, the profiles display larger values in the vertical velocities. However, the radial velocities seem to be little affected.

When the vertical velocity profiles are scaled by the maximum velocity at the orifice, one can consider changes in the shape of the profiles. We observed that the profiles far from the orifice (one silo diameter above the base, or four orifice diameters above the base) remain flat as in a free discharge. Likewise, the scaled profile at the plane of the orifice conserves the shape of the one observed for a free discharge. However, at a plane one orifice radius above the base, the vertical velocity profile becomes flatter as the overweight is increased.  {These results suggest that the use of an overweight helps to develop a mass flow rate profile even for these flat-bottomed silos, which may be of particular interest in the industry.}

{A number of open questions remain with respect to forced flows, such as: (i) Is there a simple way to upgrade the Beverloo rule to a height-dependent expression that takes into account the effect of an overweight? (ii) Is the 5/2 power in the Beverloo rule still valid? (iii) Since the acceleration of the flow rate depends on the material, what material properties are more relevant? (iv) Is the pressure profile inside the silo dramatically different from the unforced discharge? (v) What is the undelying mechanism that induce the accelerated phase to start logarithmically sooner as the overweight increases? (vi) Are there connections with related experiments where an overweight has induced drastic changes in the granular dynamics \cite{metcalfe2002}? These, and other questions, need to be addressed. The new knowledge that we earn regarding forced discharges may help develop new models to understand even the apparently simpler case of a free discharge.}

\textbf{Acknowledgments} We thank J. R. Darias for valuable discussions. This work has been supported by ANPCyT (Argentina) through grant PICT 2012-2155 and UTN (Argentina) through grant PID MAUTNLP-2184.

\end{document}